\documentclass[12pt]{article}
\usepackage[utf8]{inputenc}
\usepackage{lineno}
\usepackage{comment}
\usepackage{siunitx}
\usepackage{subcaption}
\usepackage{authblk} 
\usepackage[style=numeric,sorting=none]{biblatex}
\usepackage{graphicx}
\usepackage{amsmath}
\usepackage[english]{babel}
\usepackage{braket}

\begin{document}

\title{Estimating the electrical energy cost of performing arbitrary state preparation using qubits and qudits in integrated photonic circuits}

\author[1]{Maria Carolina Volpato}
\author[1]{Gabriel da Silva Sampaio}
\author[1,*]{Pierre-Louis de Assis}
\affil[1]{\textit{"Gleb Wataghin" Institute of Physics, Universidade Estadual de Campinas, Campinas, Brazil.}}
\affil[*]{Corresponding author: plouis@unicamp.br}

\date{} 

\maketitle

\begin{abstract}
As quantum photonic hardware scales toward computationally relevant sizes, energy consumption has emerged as a key constraint. Programmable photonic integrated circuits, composed of interferometer meshes with tunable phase modulators, provide a flexible platform for quantum information processing using both qubits and qudits. In this work, we analyze the energetic cost of such devices by focusing on arbitrary quantum state preparation, a resource-intensive task central to quantum simulation and information processing. Using a common hardware, we benchmark qudit-based implementations, gate-based quantum computation, and measurement-based quantum computation. We find that while qudit encodings are attractive at small scale, their footprint and reconfiguration costs grow rapidly with system size, whereas qubit-based approaches incur significant overhead from entangling operations, feedforward, and reprogramming. Across all paradigms, scaling beyond a few tens of qubits renders either the energy consumption or the total preparation time prohibitive on fully programmable PICs. Our results highlight the need for optimized, task-specific photonic architectures to enable energy-efficient scaling.
\end{abstract}

\section{Introduction}
Since the seminal proposals of Knill, Laflamme, and Milburn in 2000~\cite{knill2000efficient,knill2000thresholds} and 2001~\cite{Knill2001}, discrete-variable photonic quantum information processing has progressed from table-top optical setups to scalable photonic integrated circuits (PICs). This transition accelerated as high-quality lithography equipment and photonic foundry services became available and allowed an increase in device size, complexity, and fabrication throughput \cite{Wang2020}.

In the past decade, significant progress has been made in gate-based (GBQC) \cite{lemr2011preparation,lu2007demonstration,maring2024versatile}, measurement-based (MBQC) \cite{Raussendorf2001,nielsen2004optical,briegel2009measurement}, and more recently fusion-based quantum computing (FBQC) architectures \cite{rudolph2017optimistic,bartolucci2023fusion} for quantum devices for quantum information processing (QIP), quantum state preparation (QSP), and quantum simulation (QS) tasks \cite{matthews2009manipulation,shadbolt2012generating,shen2017deep,bogaerts2020programmable,humphreys2013quantum,amiri2023comparing}. Despite these advances, the scale remains relatively small compared to the hundreds of qubits now available on other platforms. Similar progress has been achieved with devices utilizing continuous-variable and hybrid encodings of quantum information \cite{madsen2022quantum, wang2018integrated}.

Across all these approaches, scalability is ultimately constrained by the underlying programmable photonic hardware. At the core of quantum integrated photonic devices lie programmable Mach-Zehnder Interferometers (MZIs). By incorporating two phase modulators, a balanced MZI can act as a tunable beam splitter. This tunability, which is widely used in classical telecommunications, is not only important to allow for reconfigurable circuits, but also to compensate for fabrication imperfections. By concatenating two MZIs it is even possible to build an ideal tunable beam splitter \cite{Miller2015}. Integrated arrays of interferometers are also being investigated for classical applications such as machine learning and signal processing \cite{shen2017deep,shastri2021photonics}, which has contributed to the development of tuning algorithms and the exploration of more complex architectures.

To address computationally relevant tasks, photonic circuits must scale to the preparation of states involving $\mathcal{O}(10^2)$–$\mathcal{O}(10^3)$ qubits \cite{Preskill2018quantumcomputingin}. As circuit size grows, energy consumption becomes a key bottleneck \cite{PRXQuantum.3.020101,Parker2023,Fellous-Asiani2023}. This motivates our central question: what is the energy cost of programming large-scale PICs composed of MZI arrays using electro-optical modulators (EOMs)? Although thermal phase shifters dominate current PIC demonstrations, they dissipate energy continuously. EOMs, by contrast, act as capacitive elements. They are nearly lossless at low operating frequencies but become dissipative in the \si{GHz} regime~\cite{miller2012energy}, precisely the operating range compatible with typical single-photon generation rates \cite{tomm2021bright}. 

To perform a quantitative evaluation of the energy consumption of quantum PICs, we constrain our analysis to the task of arbitrary QSP (aQSP). We have chosen this task because it allows for a more general treatment, as opposed to QIP, which would require selecting one or more algorithms for comparison, and encoding information on a quantum state is a preliminary task to QIP and is a core task of variational QS algorithms such as variational quantum eigensolvers (VQE) \cite{Blunt2022}. While previous studies have focused on estimating the energy consumption associated with photonic quantum platforms at the system level—including sources, detectors, and classical processing—our analysis targets the energetic cost of realizing quantum operations within the photonic circuit. By isolating the energy required to program and reconfigure interferometric networks for aQSP, we provide a complementary perspective on scalability limits in integrated quantum photonics \cite{soret2026quantum}.

We investigated the energetic cost of implementing path-encoded qubits and qudits within GBQC and MBQC. To ensure a fair comparison, our analysis adopts a common hardware baseline based on programmable integrated circuits fabricated in thin-film lithium niobate (TFLN), a platform that combines low-loss photonic integration with high-speed electro-optic modulation in the \si{GHz} regime \cite{zheng2023electro,wang2018integrated,hu2025integrated}. 

Even though qudits require exponentially more waveguides to implement a state of a given dimension than qubits, we use them as a benchmark. Not only are \textit{d}-level systems much simpler to implement and manipulate using integrated photonics than other quantum platforms, but decomposition algorithms exist for square meshes of MZIs to implement any arbitrary unitary with a dimension defined by the number of ports of the array \cite{cybenko2001reducing,clements2016optimal}. The preparation of an arbitrary multi-qubit quantum state in the gate-based model requires a number of Controlled-NOT (CNOT) gates that scales exponentially with the number of qubits $n$ \cite{PhysRevA.52.3457}. Here, we derive an upper bound on the required number of CNOT gates assuming only nearest-neighbor interactions, thereby avoiding waveguide crossings, which can otherwise introduce significant insertion losses in photonic circuits. 

A further crucial consideration is that linear-optical CNOT gates are inherently probabilistic. To address this limitation, we investigate the near-deterministic CNOT gate proposed within the KLM framework \cite{Knill2001,knill2000efficient}, which increases the gate success probability from $1/4$ to $n^2/(n+1)^2$ by employing additional photons and optical operations. In this scheme, multiple probabilistic gates are prepared offline in parallel using $n$-photon entangled ancilla states, and a control signal is routed to the output mode upon successful heralding. As $n$ increases, the success probability asymptotically approaches unity, albeit at the cost of substantially increased resource and energy overhead. Since GBQC and MBQC are computationally equivalent \cite{Raussendorf2001}, this analysis also enables an estimation of the energetic cost associated with MBQC implementations on photonic integrated circuits.

To mitigate the energetic overhead associated with aQSP, alternative strategies can be adopted, such as time-domain demultiplexing of qubits \cite{pegoraro2024demonstration}, which allows a single physical CNOT gate to be reused sequentially for the computation. Finally, we briefly comment on FBQC as a fault-tolerant photonic paradigm, emphasizing its role as a theoretical benchmark while noting the substantial practical challenges that currently hinder its implementation on integrated photonic platforms.

\section{Standard hardware}\label{sec:hardware}
\begin{figure}[h!]
        \centering
        \includegraphics[scale=0.5]{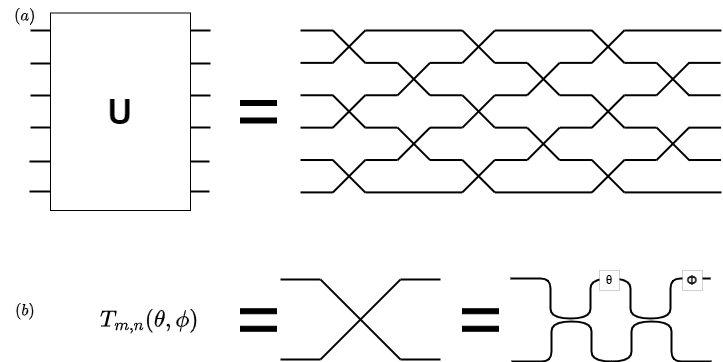}
        \caption{(a) Universal n-mode multiport interferometer (shown here for $n=6$) can be implemented using a mesh of $N=n(n-1)/2$ beam splitters demonstrated in Ref.\cite{clements2016optimal}; (b) The crossing represents a variable beam splitter described by $T_{m,n}(\theta,\phi)$ matrix, which can be implemented by a MZI consisting of two $50:50$ directional couplers, followed by a phase shifter at one input port.}
        \label{fig:clements}
    \end{figure}

In order to have a proper basis of comparison between the qubit and qudit aQSP, we will consider that both are performed using the same standard hardware, based on a programmable array of Mach-Zehnder Interferometers (MZIs) in a topology proposed by Clements \textit{et al.} \cite{clements2016optimal}, illustrated in Fig. \ref{fig:clements}. Here we consider the hardware as programmable if its configuration can be altered dynamically, resulting in a different output for the same input. This is more similar to the electronic FPGA circuits than to programmable computers in the sense of those which use a von Neumann architecture. 

Given the time and cost associated with the design and fabrication of photonic integrated circuits, full reconfigurability is an intuitively attractive feature. We therefore focus on a fully programmable interferometric architecture, as it enables the direct implementation of aQSP in qudit systems through unitary transformations, as illustrated in Fig.~\ref{fig:clements}(a) and discussed in Sec.~\ref{subsec:qudits}. The same architecture can also be employed to implement entangling operations, such as CNOT gates, required for gate-based and measurement-based quantum computation, as detailed in Secs.~\ref{subsec:qubits} and~\ref{subsec:mbqc}. In addition, single-qubit operations can be realized using individual Mach–Zehnder interferometers.

The energetic cost of such reconfigurability is dominated by the repeated programming of the phase shifters that define each interferometer. EOMs can be modeled as capacitive elements, and under this assumption the energy required to program a photonic circuit containing $N$ Mach–Zehnder interferometers MZIs can be expressed as

\begin{align}
    E(N)= \frac{C \times \sum ^{2N} V(\varphi)^2}{2},
    \label{energy_c}
\end{align}
where $C$ is the capacitance of a single EOM and $V(\varphi)=V_\pi,\varphi/\pi$ is the voltage required to induce a phase shift $\varphi$. Each MZI requires two independent phase modulators, resulting in a total of $2N$ EOMs, as illustrated in Fig.~\ref{fig:clements}(b).

\section{State preparation} \label{sec:QSP}
We adopt arbitrary quantum state preparation (aQSP) as the benchmark task to compare the energy consumption of integrated photonic circuits operating with qubits and qudits. In both cases, quantum information is encoded in the path degree of freedom of single photons, and arbitrary operations are implemented using MZIs acting as programmable beam splitters \cite{politi2008silica}. We focus on aQSP because it is a pre-requisite for QIP tasks, since the input state needs to be encoded in a quantum system prior to processing. The task of preparing a state consists of starting from an initial ``blank state'' input, and reaching the target state through a series of unitary operations on that input, which are defined by the choice of target state.

\subsection{Qubits} \label{subsec:qudits}
An arbitrary $n$-dimensional path-encoded quantum state can be programmed by sending a single photon through a programmable interferometer with $n$ input and $n$ output ports. These interferometers are typically implemented as meshes of beam splitters and phase shifters \cite{reck1994experimental,clements2016optimal}, which aligns with the standard hardware considered in this work. Cybenko \cite{cybenko2001reducing} and, later, Clements \textit{et al.} \cite{clements2016optimal} demonstrated that any unitary matrix can be decomposed as
    \begin{align}
        U=D\left(\prod_{m,n}T_{m,n}(\theta, \phi) \right),
    \end{align}
where $D$ is a diagonal matrix with complex elements with a modulus equal to one on the diagonal, and each $T_{m,n}$ matrix corresponds to a lossless beam splitter between channels $m$ and $n$ $(m=n-1)$ with reflectivity $\cos{\theta}$ $(\theta \in [0,\pi/2))$ and a phase shift $\phi$ $(\theta \in [0,2\pi))$ at input $m$. 
    
Based on this approach, we can identify the appropriate unitary matrix that prepares the any desired state when starting from $\ket{0}$, which corresponds to sending the photon through the first input waveguide of the interferometer.
    
\subsection{Gate-based quantum computation (GBQC)} \label{subsec:qubits}
To prepare an n-dimensional state using qubits, $2\log_2(n)$ waveguides and $\log_2(n)$ photons are necessary. CNOT gates are essential for aQSP when considering an initial state $|0\rangle ^{\otimes n}$, since the preparation circuit must be able to generate entanglement \cite{barenco1995elementary}. We have chosen a separable initial state so as to avoid masking energy costs by assuming a highly-entangled resource whose energetic balance would not be treated in our analysis.
    
CNOT gates are well known to be non-deterministic in linear optical quantum information processing (LOQIP). The CNOT gate as first proposed in the context of fully programmable photonic circuits using the Clements topology \cite{o2003demonstration} has a success probability of $1/9$ and is self-heralded. While this is attractive because it does not require ancillary photons, the approach is not scalable: implementing a multi-gate circuit requires cascading multiple CNOT operations, causing the overall success probability to decrease exponentially as $(1/9)^{N_{\mathrm{CNOT}}}$. Moreover, verifying correct operation still requires measurement of the output state.
    
Another proposal is to use the KLM scheme \cite{knill2000efficient,Knill2001}, which, by leveraging an entangled initial state, auxiliary photons, post-selection and feedforward, effectively increases the probability of success in implementing CNOT gates, achieving a near-deterministic gate. This capability is crucial for aQSP, as the generation of entanglement through CNOT gates is a fundamental requirement. However, the scheme incurs significant resource overhead—including additional photons and classical feedforward control—highlighting the trade-off between deterministic operation and resource efficiency. We note that our analysis does not consider how the resource state $\ket{t_n}$ is generated, which in principle could be prepared offline using CNOTs and then injected into the circuit.

In fully programmable photonic circuits, such as those based on the Clements topology, implementing CNOT gates within the KLM framework imposes substantial scaling requirements on the interferometric network. Even the basic KLM CNOT, which succeeds with probability $1/4$, requires a large interferometer to realize the corresponding unitary transformation, typically implemented as a $24 \times 24$ array of Mach–Zehnder interferometers. Increasing the success probability toward near-deterministic operation further demands the use of additional ancilla photons and teleportation-based gate constructions. As shown in our analysis, these added resources translate directly into a significant increase in the energy required to program and operate the circuit. Consequently, at the algorithmic level, the energetic cost of aQSP in fully programmable photonic platforms is largely governed by the number of entangling operations required.
    
\begin{figure}[h!]
        \centering
        \includegraphics[trim={3cm, 24.5cm, 12cm, 1.7cm},clip, width=0.6\textwidth]{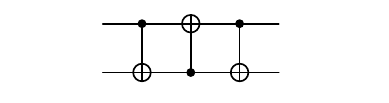}
        \caption{Circuit representation of a two-qubit state preparation using three consecutive CNOT gates. Fifteen single-qubit operations (SQOs) are also required \cite{vatan2004optimal}, but are omitted here for clarity.}
        \label{fig:2qubit}
        \end{figure}

Therefore, in order to optimize the energetic cost of operation, we seek to minimize the number of CNOT gates \cite{divincenzo1994results} used for aQSP. To prepare an arbitrary state of 2 qubits, a total of 3 CNOTs and 15 SQOs (omitted for clarity) are required \cite{vatan2004optimal}, as shown in Fig. \ref{fig:2qubit}.
    
However, as the number of qubits increases, finding the optimal circuit becomes more complex, which has led to the development of multiple algorithms aimed at minimizing the number of CNOTs and SQOs \cite{shende2004minimal,bergholm2005quantum,plesch2011quantum}. Bergholm \textit{et al.} \cite{bergholm2005quantum} established an upper bound on the number of CNOTs required for state preparation using only CNOTs and SQOs in nearest-neighbor configurations, given by

    \begin{align}
        C_B(n)=\frac{10}{3} 2^n + 2n^2-12n+\left\{\begin{matrix}
\frac{14}{3} \, \text{, n even.}\\ \\
\frac{10}{3} \, \text{, n odd.}
\end{matrix}\right.
    \end{align} 
        
Plesch and Bruckner \cite{plesch2011quantum} presented a more refined upper bound for the number of CNOTs, by dividing the state preparation in four phases where the last two can be performed in parallel. The new bound for the number of CNOTs is given by 
    
    \begin{equation}
    C_{P}(n)=\left\{\begin{matrix}\frac{1}{48}\left(80-21 \cdot 2^{2+n/2}+115\cdot2^n \right) \text{, n even.}\\ \\
    \frac{1}{48}\left(80+23 \cdot 2^{n}-15\cdot2^{(7+n)/2} \right)\text{, n odd.}
    \end{matrix}\right.
    \end{equation}
    
\begin{figure}[h!]
        \centering
        \includegraphics[trim={12.5cm 24cm 3cm 1.7cm},clip, width=0.6\textwidth]{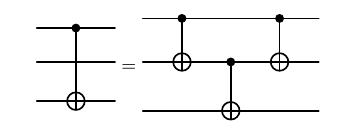}
        \caption{Representation of the decomposition of a long-range CNOT gate into a sequence of nearest-neighbor CNOTs \cite{bergholm2005quantum}.}
        \label{fig:cnot1_3}
        \end{figure}

Their proposal, however, makes use of long-range CNOTs, and must be adapted in order to satisfy the requirements imposed by integrated photonics. We do so by using the decomposition of a long-range CNOT into a sequence of nearest-neighbor CNOTs proposed in \cite{bergholm2005quantum}, shown in Fig. \ref{fig:cnot1_3}.
    
\begin{figure}[h!]
        \centering
        \includegraphics[trim={2cm 16.6cm 11cm 1.7cm},clip,width=0.6\textwidth]{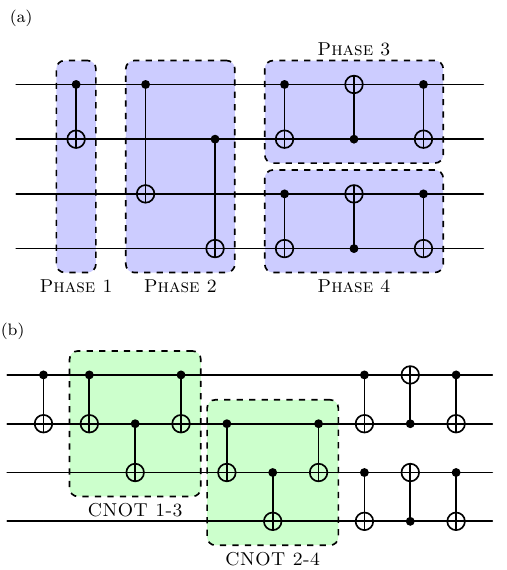}
        \caption{Gate sequence for preparation of an arbitrary four-qubit state using (a) the Plesch circuit for QSP \cite{plesch2011quantum}, with the four steps highlighted in blue boxes; (b) a modified Plesch circuit which only uses nearest-neighbor CNOTs, with the CNOT decompositions highlighted in green boxes for clarity.}
        \label{fig:caslav_mod}
        \end{figure}

The upper bound $C_m(n)$ for the number of CNOTs when modifying the Plesch decomposition to only use nearest-neighbor gates is given by
    \begin{equation}
        C_{m}(n)= \left\{\begin{matrix}
C_P(n)+n \text{, n even,}\\ \\
C_P(n)+n-1 \text{, n odd,}
\end{matrix}\right.
    \end{equation} 
also considering $n$ as the number of qubits. As an example we show in Fig. \ref{fig:caslav_mod} the circuit for 4 qubits using the standard Plesch circuit (a) and our modified version (b), where the decompositions of long-range CNOTs are highlighted in green.

\begin{figure}[h!]
        \centering
        \includegraphics[width=0.7\textwidth]{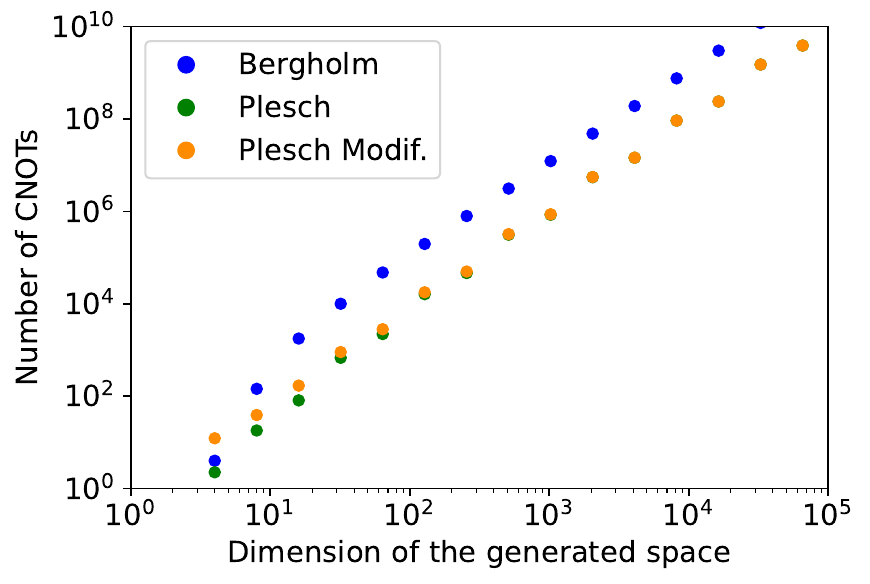}
        \caption{Comparison of the number of CNOT gates required for qubit state preparation using different methods: Bergholm method (blue), original Plesch method (green), and a modified Plesch method restricted to nearest-neighbor CNOTs (yellow). The modified circuit highlights CNOT decompositions for clarity. As shown, while the Bergholm method is more efficient for $n=2$ qubits, the adapted Plesch method becomes more efficient as $n$ increases, closely approaching the upper bound.}
        \label{fig:number}
        \end{figure}

As can be seen in Fig. \ref{fig:number}, the Bergholm method is more efficient for $n=2$ qubits, but as $n$ increases, our adaptation of the Plesch method requires less CNOTs, staying close to the original bound, even considering only nearest-neighbor CNOTs.

The restriction to nearest-neighbor CNOTs is particularly relevant for integrated photonic implementations, where interactions between distant qubits require waveguide crossings. While such crossings are feasible in integrated photonics, they introduce significant insertion losses, which can degrade circuit performance \cite{celo2014low,he2025broadband,zhang2024300}. Also, we adopt this upper bound because it is the only result in the literature that explicitly accounts for the SQOs, which are essential for encoding arbitrary quantum states. To best of our knowledge, the only lower-bound result for CNOT gates in this context considers just the two-qubit case \cite{shende2004minimal}, and it is not clear whether the same decomposition can be extended to larger systems with nearest-neighbor constraints.

\subsection{Measurement-based quantum computation (MBQC)} \label{subsec:mbqc}
Measurement-based quantum computation (MBQC) provides an alternative paradigm to the circuit model, in which quantum information processing is carried out through local measurements on a highly entangled resource state, most commonly a cluster state \cite{Raussendorf2001,briegel2009measurement}. In contrast to the standard circuit model, where quantum information is manipulated via coherent quantum gates, MBQC replaces dynamical gate execution with sequences of adaptive measurements whose bases depend on prior measurement outcomes.

In this framework, computation proceeds in two conceptually distinct stages. First, the system is initialized in a highly entangled multipartite state, such as a 2D cluster state, prepared independently of the specific algorithm to be executed. This state therefore acts as a universal computational resource. In the second stage, individual qubits are measured sequentially in appropriately chosen bases, and the classical outcomes determine both the effective logical operations and the choice of subsequent measurement bases. Logical single-qubit rotations, Pauli $\sigma_x$ and $\sigma_z$ corrections, and multi-qubit entangling operations such as CNOT gates are realized through the geometry of the cluster state and the adaptive measurement pattern.

The computational equivalence between MBQC and GBQC enables a direct mapping between measurement sequences and effective gate operations \cite{Raussendorf2001}. In the context of photonic integrated circuits, this equivalence allows the resource and energy costs of MBQC to be benchmarked by translating measurement patterns into the corresponding number of single-qubit rotations and entangling gates. 

\section{Evaluation of the energy cost}
The energy consumption is evaluated by calculating the average electrical energy required to program the standard hardware defined in Section~\ref{sec:hardware}, assuming that each MZI is tuned using state-of-the-art lithium niobate electro-optic modulators (EOMs) \cite{wang2018integrated}. We consider EOMs with a length of \SI{5}{mm}, characterized by a half-wave voltage $V_\pi=$ \SI{4.4}{V} and a capacitance $C=$ \SI{0.03}{pF}. In all cases, we impose a maximum state-preparation time of one day, and a photon generation of \SI{1}{GHz} \cite{senellart2017high} which sets the operational repetition rate and ensures a consistent comparison between different architectures and decomposition strategies.

\subsection{Qudits}
To evaluate the energy cost of preparing qudit states, we generated $10^4$ Haar-random unitary matrices for dimensions ranging from 2 to 64 and calculated the corresponding phases required to implement each unitary. We then performed a quadratic fit of the average energy cost as a function of dimension, which allows extrapolation to arbitrary dimensions. As shown in Fig.~\ref{fig:results} in blue, the expected exponential scaling of energy consumption with state dimension is evident. For a state of dimension $2^{50}$ (corresponding to 50 qubits), the required energy approaches the total electricity generation of the U.S. in a whole year \cite{EIA_US_Generation_2024}.

\subsection{GBQC}
\begin{figure}[h!]
        \centering
        \includegraphics[width=0.6\textwidth]{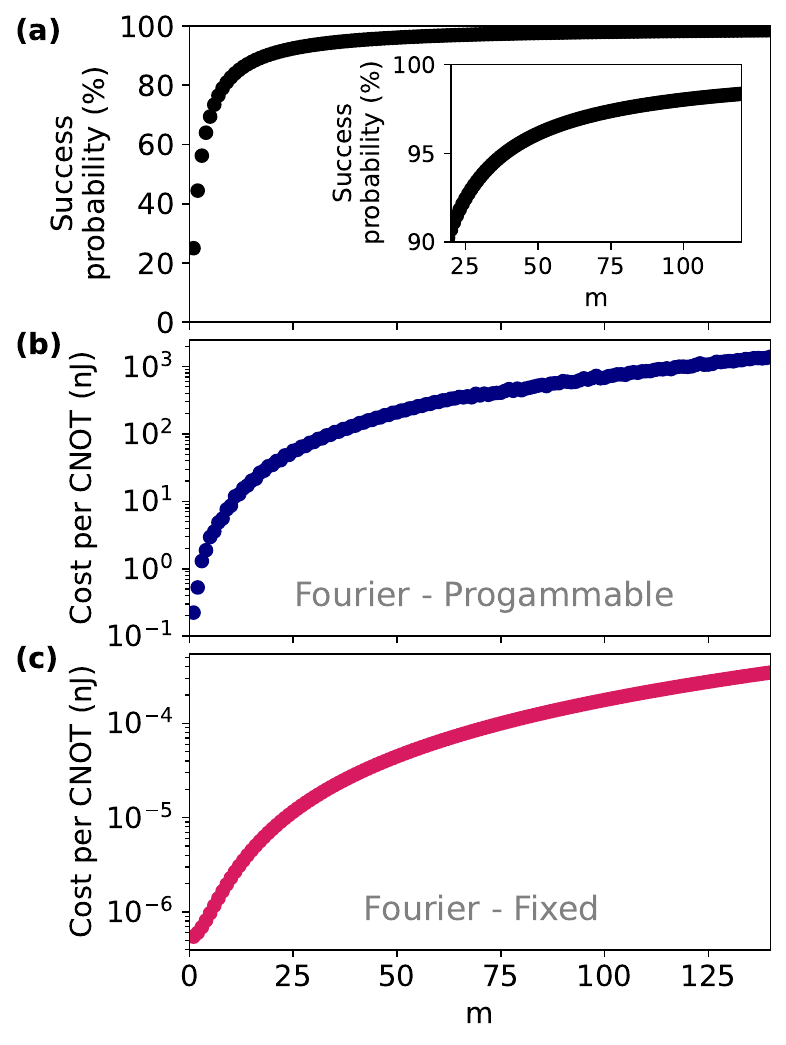}
        \caption{(a) Success probability of a CNOT gate in the near-deterministic KLM framework~\cite{Knill2001}, given by $P_{\text{CNOT}}=m^2/(m+1)^2$, where $m$ is the dimension of the entangled ancilla state $\ket{t_m}$. Energy cost to implement a single CNOT gate as a function of $m$, considering the $\hat{F_m}$ beeing implemented by (b) a fully programmable MZI array and (c) a fixed photonic circuit with \SI{1}{\percent} correction~\cite{bandyopadhyay2021hardware}.}
        \label{fig:KLM}
        \end{figure}

For GBQC state preparation, we consider the KLM framework as the only known linear-optical approach enabling near-deterministic CNOT gates \cite{knill2000efficient,Knill2001}. Fig.~\ref{fig:KLM}(a) shows how the success probability of a linear-optical CNOT gate is enhanced via teleportation using an entangled ancilla state $\ket{t_m}$ whose dimension $m$ determines the gate success probability $P_{\text{CNOT}}=m^2/(m+1)^2$. While increasing $m$ allows the CNOT operation to asymptotically approach determinism, it also leads to a substantial increase in the resources required to prepare the corresponding ancilla state. 

Fig. \ref{fig:KLM}(b) shows the estimated energy cost associated with implementing a single near-deterministic CNOT gate, where the Fourier transform $\hat{F_m}$, which is a necessary step to boost to near deterministic CNOTs, is synthesized using a Clements decomposition. In this fully programmable approach, $\hat{F_m}$ is implemented by an $m \times m$ Mach–Zehnder interferometer (MZI) array comprising all required beam splitters and phase shifters, as well as additional trimming operations to compensate for fabrication imperfections.

One way to significantly reduce the energy consumption is to replace fully programmable interferometers with fixed, application-specific photonic circuits, where a residual trimming of only \SI{1}{\percent} is assumed to correct for fabrication defects\cite{bandyopadhyay2021hardware}. As shown in Fig.~\ref{fig:KLM}(c), this strategy reduces the energy required to implement a single CNOT gate by approximately five orders of magnitude compared to the fully programmable case. 

To estimate the total energetic cost of universal quantum state preparation, we compute the cumulative energy required to implement the complete set of CNOT gates necessary for aQSP, including the associated SQOs, shown in green in Fig.~\ref{fig:results}. Even under optimistic assumptions, this fully spatially multiplexed approach remains energetically prohibitive when a large number of CNOT gates is required.

An alternative strategy to mitigate this cost is to demultiplex qubits in time \cite{pegoraro2024demonstration,larsen2019deterministic}, allowing a single near-deterministic CNOT gate to be reused sequentially throughout the computation. This approach requires the use of photonic memory elements \cite{lvovsky2009optical,reim2010towards} or programmable delay lines \cite{bose2024anneal,song2023electro}. To account for their contribution, we assume that the energy cost of a single delay operation is equivalent to that of four Mach–Zehnder interferometers operated in a cross-to-bar configuration, and that such a delay is required for each CNOT operation. Since our decomposition restricts interactions to nearest neighbors only, the temporal spacing between qubits can be taken to be uniform, enabling a regular and hardware-efficient timing architecture.

    \begin{figure}[h!]
        \centering
        \includegraphics[width=0.6\textwidth]{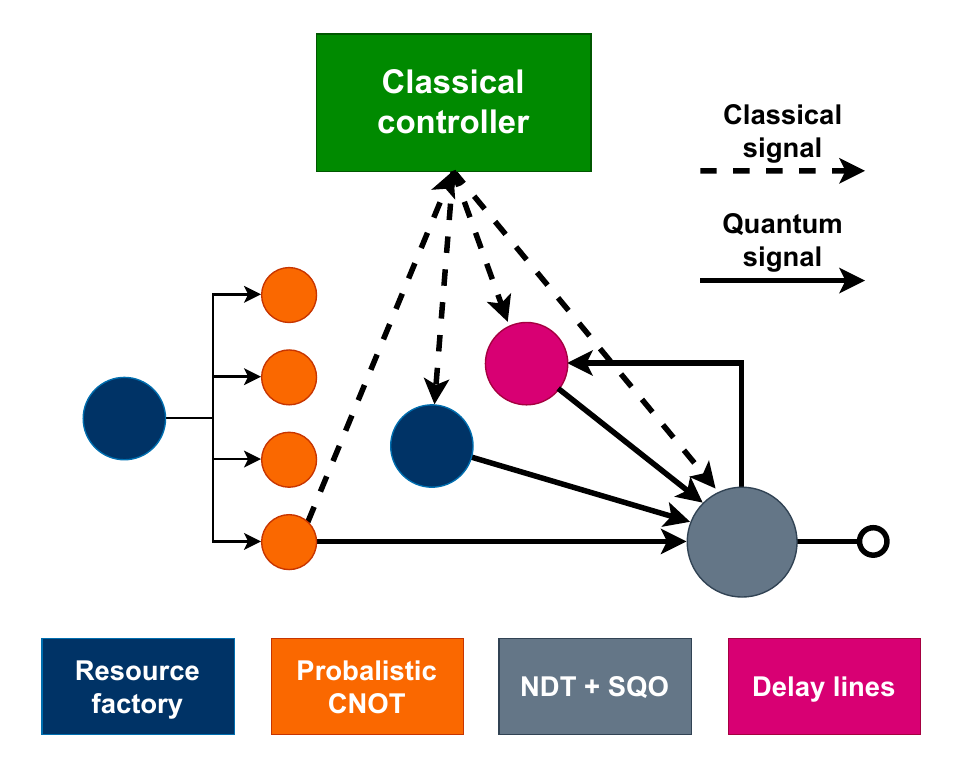}
        \caption{Schematic of a time-demultiplexed GBQC protocol based on near-deterministic KLM CNOT gates. A resource factory generates photons that feed probabilistic CNOT gates. Successful gate operations are heralded and communicated to a classical controller, which triggers the corresponding near-deterministic teleportation (NDT) steps and single-quantum operations (SQO). Quantum states are stored in delay lines or photonic memories until released for further processing. This architecture enables the reuse of a single near-deterministic CNOT gate, substantially reducing the overall energetic cost of arbitrary quantum state preparation.}
        \label{fig:protocol}
        \end{figure}

Fig. \ref{fig:protocol} illustrates a protocol based on this time-demultiplexed architecture. A resource factory continuously generates photons to supply the photonic circuit, which feeds probabilistic CNOT gates. Upon successful operation, a heralding signal is sent to the classical controller, which triggers the preparation of the near-deterministic teleportation steps and the associated SQOs. The resulting quantum states are stored in a photonic memory until the classical controller releases them for subsequent NDT+SQO operations. This cycle is repeated until the target quantum state is fully prepared. The corresponding energy estimates for this time-demultiplexed approach are reported in purple in Fig.~\ref{fig:results}, highlighting its favorable scaling compared to fully spatially multiplexed architectures.

\subsection{MBQC}
\begin{figure}[h!]
        \centering
        \includegraphics[width=0.8\textwidth]{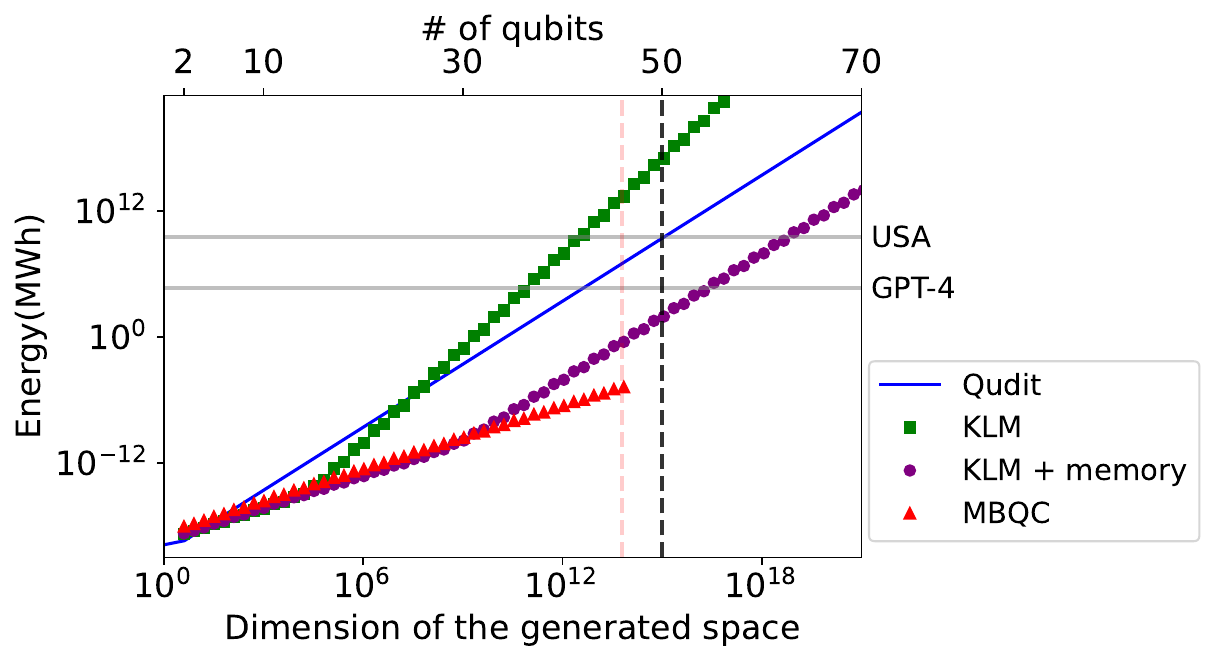}
        \caption{Estimated energy cost (in \si{\mega\watt}) for universal quantum state preparation under a fixed target preparation time of one day, comparing different decomposition strategies: qudit-based encoding in blue, GBQC via the KLM scheme using fully programmable circuits for the teleportation step in green, GBQC with pre-implemented hardware combined with photonic memory in purple, and MBQC in red. The pink dashed line indicates the maximum Hilbert-space dimension accessible within MBQC under the one-day preparation constraint. Gray horizontal lines show, for comparison, the total U.S. electricity generation in 2024 \cite{EIA_US_Generation_2024} and the estimated energy consumption required to train the GPT-4 model \cite{groes2023carbon}.}
        \label{fig:results}
        \end{figure}

For the implementation of MBQC, we exploit its equivalence with gate-based quantum computation to estimate the energy costs associated with CNOT gates and arbitrary single-qubit rotations \cite{Raussendorf2001}, using a mapping to Mach–Zehnder interferometer meshes \cite{kwon2024quantum}. Although MBQC relies on adaptive measurements, we consider the average cost required to implement aQSP, including projections onto arbitrary angles in the $X$ and $Z$ bases. Pauli corrections ($\sigma_x$ and $\sigma_z$) are assumed cost-free, with only the trimming of MZIs contributing to the energy budget. As controlled-Z (CZ) gates are probabilistic, we multiply the effective number of CNOTs by three to account for resource overhead. As shown in Fig.~\ref{fig:results} in red, the energy cost for preparing an arbitrary quantum state using MBQC is slightly lower than that of gate-based approaches, yet still exhibits exponential scaling due to the required number of entangling gates and single-qubit rotations. Finally, by imposing a maximum state-preparation time of one day, we find that only Hilbert-space dimensions up to $2^{46}$ are accessible within this framework, a constraint indicated by the pink dashed line in Fig.~\ref{fig:results}. We emphasize that the energy cost associated with generating the cluster state itself is not included in this estimate.

Implementations based on qudits remain attractive at low dimensionality, as encoding a single photon across multiple spatial paths is experimentally feasible. However, as the system size increases, practical limitations rapidly emerge. For example, even assuming compact electro-optic modulators with a length of \SI{500}{\micro\meter}, an order of magnitude smaller than the \SI{5}{\milli\meter} devices used in our energy estimates implementing, an 8-qubit state would already require an area comparable to that of a full 12-inch wafer. This highlights that, despite their conceptual simplicity, path-encoded qudit architectures face severe footprint and scalability constraints.

In contrast, qubit-based photonic computation can, in principle, adopt a modular architecture. Within GBQC, KLM-style CNOT gates can approach near-deterministic operation provided that the required entangled resource states are prepared offline using additional CNOT gates. While offline preparation relaxes the need to generate these resources in real time, it inherently limits the number of computational steps that can be executed sequentially. Furthermore, photonic qubits must be stored between operations and demultiplexing, necessitating long, programmable delay lines or quantum memories, which introduce substantial routing overhead and architectural complexity.

MBQC offers an alternative paradigm in which computation proceeds via adaptive single-qubit measurements performed on a pre-prepared cluster state. In this approach, measurement outcomes must be processed and fed forward in real time, requiring classical control, detection, and reconfiguration to operate at the single-photon emission rate. As the number of qubits increases, the total number of adaptive operations grows accordingly, eventually imposing a practical time constraint. In particular, for arbitrary quantum state preparation, this scaling prevents reaching the NISQ regime ($\approx$50 qubits) within a one-day preparation window.

As expected, implementing arbitrary quantum state preparation on fully programmable photonic integrated circuits remains energetically costly, even when relying on state-of-the-art electro-optic modulators. However, since EOMs behave as capacitive elements, part of the energy used to reprogram the circuit can, in principle, be recovered when the programming rate is slower than the characteristic $RC$ time of the modulator, as is typically the case for qudit-based implementations. More specialized photonic circuits designed specifically to this task can be significantly more energy efficient, as they require only small adjustments to beam-splitter settings rather than full circuit reprogramming. Nonetheless, reaching the NISQ limit with photons remains a practical hardware challenge, highlighting the need for optimized device designs and innovative circuit architectures.

\section{Benchmark against classical computing}
Beyond state preparation, cryptographically relevant algorithms further illustrate the energetic implications of scalable quantum computation. Shor’s algorithm, for example, requires a large number of logical operations for practical problem sizes \cite{shor1999polynomial}. Factoring a 3072-bit RSA key is estimated to require on the order of $6 \times 10^3$ logical qubits and $10^{13}$ Toffoli gates \cite{roetteler2017quantum}. Using our energy estimates for fully programmable GBQC corresponds to a minimum energy cost of approximately \SI{2}{\mega\watt\hour}, while for an MBQC implementation, the corresponding energy cost to roughly \SI{1e-5}{\mega\watt\hour}. Importantly, this computational task is not feasible for classical computers, as the time and energy required to factor a 3072-bit RSA key using state-of-the-art classical algorithms would exceed any realistic computational and energetic resources.

\section{A comment on FBQC}
Finally, we briefly comment on Fusion-Based Quantum Computation (FBQC), a theoretical approach that has gained attention for photonic fault-tolerant architectures \cite{kok2007linear,browne2005resource,rudolph2017optimistic}. FBQC is designed to implement surface-code logical qubits by combining small entangled resource states through probabilistic fusion measurements \cite{bartolucci2023fusion}. While the model provides a promising framework for error-corrected photonic quantum computation, to date there is no fully integrated photonic implementation, and realizing it would require extremely large numbers of high-quality single photons, low-loss routing, and fast buffering or quantum memories \cite{bartolucci2023fusion,meng2025temporal}. Consequently, FBQC currently serves mainly as a theoretical benchmark for resource and energetic scaling rather than a near-term experimental platform.

\section{Discussion about arbitrary QSP}
Barenco \textit{et al.} \cite{PhysRevA.52.3457} argued that most QSP realizations may not be “computationally interesting” because they require exponentially many operations to implement. This is a good pragmatic argument since it is clear that any quantum advantage is lost if exponentially increasing costs incur, and favors the use of programmable circuits. It should be possible to use them to efficiently prepare computationally interesting states, as opposed to attacking the arbitrary state preparation task. 

This assumption becomes even more reasonable when considering that some QIP applications will leverage the higher efficiency of quantum computers to process classical inputs into classical outputs. In such cases, the QSP step encoding the initial information for QIP does not require a circuit capable of preparing entangled states. For gate-based devices, as in our study, entanglement arises during the computation. However, evaluating the energy consumption of different quantum algorithms falls outside the scope of this work.

The need for arbitrary QSP persists, however, for quantum simulations tasks such as determining the energy levels of molecules of interest via variational quantum eigensolver or quantum phase estimation algorithms. Both algorithms require implementing arbitrary unitary operations. For these cases, photonic implementations will require dedicated units, leaving the probabilistic nature of the CNOT gates as the main bottleneck.

\section{Conclusions}
In this work, we investigated the energetic cost of programmable photonic integrated circuits for quantum information processing, using aQSP as a representative and resource-intensive task. By adopting a common hardware baseline based on thin-film lithium niobate technology, we enabled a consistent comparison between different computational models and encoding strategies within the same physical platform. Our analysis shows that fully programmable, qudit-based implementations can consume more energy than their qubit-based counterparts once the system exceeds a small number of qubits, even when assuming state-of-the-art electro-optic modulators. More generally, the energetic cost of arbitrary state preparation in qudits rapidly becomes prohibitive as the system size increases, due to the large number of interferometric elements that must be actively reconfigured.

In gate-based photonic quantum computation, the dominant overhead arises from the resources required to implement near-deterministic KLM CNOT gates, in particular the number of photons needed for teleportation-based schemes. A significant reduction in energy consumption can be achieved by reusing a single near-deterministic CNOT gate through time demultiplexing, at the expense of introducing photonic delay lines or quantum memories for routing and storage. We also investigated measurement-based quantum computation, which can exhibit lower average energy requirements than gate-based approaches. However, MBQC is subject to a stringent constraint on the number of operations that can be performed within a given time window, as adaptive measurements and feedforward must operate at the single-photon rate. As a result, for arbitrary state preparation, this limitation prevents reaching the NISQ regime within practical preparation times.

Finally, we briefly comment on fusion-based quantum computation as a fault-tolerant photonic paradigm, emphasizing its relevance as a theoretical benchmark while noting the substantial challenges associated with its implementation on integrated photonic platforms. Overall, our results identify energy consumption and routing constraints as key limitations in the scaling of programmable photonic quantum processors and provide quantitative guidance for the design of more energy-efficient photonic quantum architectures.

\section*{Acknowledgment}
The authors would like to thank Prof. Rafael Rabelo, Prof. Breno M. G. Teixeira and Prof. Diogo O. Soares-Pinto for the fruitful discussions during the preparation of this manuscript. The authors acknowledge funding from Coordenação de Aperfeiçoamento de Pessoal de Nível Superior - Brasil (CAPES) - Finance Code 001 and by CNPq grants 465469/2014-0 (INCT-IQ) and 409821/2022-5.

\printbibliography

\end{document}